\documentclass[%
twocolumn,
superscriptaddress,
 amsmath,amssymb,
 aps,
prl,
floatfix,
]{revtex4-1}

\usepackage{graphicx}
\usepackage{dcolumn}
\usepackage{bm}

\begin{document}
\newcommand{\van}{VO$_2$}
\newcommand{\vac}{$V_O$}


\title{Doping a Bad Metal: Origin of Suppression of Metal-Insulator Transition in Non-Stoichiometric VO$_2$}


\author{P. Ganesh}
\email{ganeshp@ornl.gov}
\affiliation{
 Center for Nanophase Materials Sciences, Oak Ridge National Laboratory, Oak Ridge, Tennessee, 37831, USA
}
\author{Frank Lechermann}%
\affiliation{I. Institut f\"ur Theoretische Physik, Universit\"at~Hamburg, 
Jungiusstr.~9, D-20355~Hamburg, Germany}
\author{Ilkka Kyl\"anp\"a\"a}
\author{Jaron Krogel}
\affiliation{
 Material Science and Technology Division, Oak Ridge National Laboratory, Oak Ridge, Tennessee, 37831, USA
}
\author{Paul R. C. Kent}
\affiliation{
 Center for Nanophase Materials Sciences, Oak Ridge National Laboratory, Oak Ridge, Tennessee, 37831, USA
}
\affiliation{Computational Sciences and Engineering Division, Oak Ridge National Laboratory, Oak Ridge, Tennessee, 37831, USA
}
\author{Olle Heinonen}
\affiliation{
 Materials Science Division, Argonne National Laboratory, Lemont, Illinois 60439, USA
}





\date{\today}

\begin{abstract}
Rutile ($\sl R$) phase \van~is a quintessential example of a strongly correlated bad-metal, which undergoes a metal-insulator transition (MIT) concomitant with a structural transition to a V-V dimerized monoclinic phase below T$_{MIT} \sim 340K$. It has been experimentally shown that one can control this transition by doping \van. In particular, doping with oxygen vacancies (\vac) has been shown to completely suppress this MIT  {\em without} any structural transition. We explain this suppression by elucidating the influence of oxygen-vacancies on the electronic-structure of the metallic $\sl R$ phase \van, explicitly treating strong electron-electron correlations using dynamical mean-field theory (DMFT) as well as  diffusion Monte Carlo (DMC) flavor of quantum Monte Carlo (QMC) techniques. We show that \vac's tend to change the V-3$d$ filling away from its nominal half-filled value, with the $e_{g}^{\pi}$ orbitals competing with the otherwise dominant $a_{1g}$ orbital. Loss of this near orbital polarization of the $a_{1g}$ orbital is associated with a weakening of electron correlations, especially along the V-V dimerization direction. This removes a charge-density wave (CDW) instability along this direction above a critical doping concentration, which further suppresses the metal-insulator transition. Our study also suggests that the MIT is predominantly driven by a correlation-induced CDW instability along the V-V dimerization direction.   

\end{abstract}

\maketitle

\begin{figure}[b]
 \includegraphics[width=3.3in]{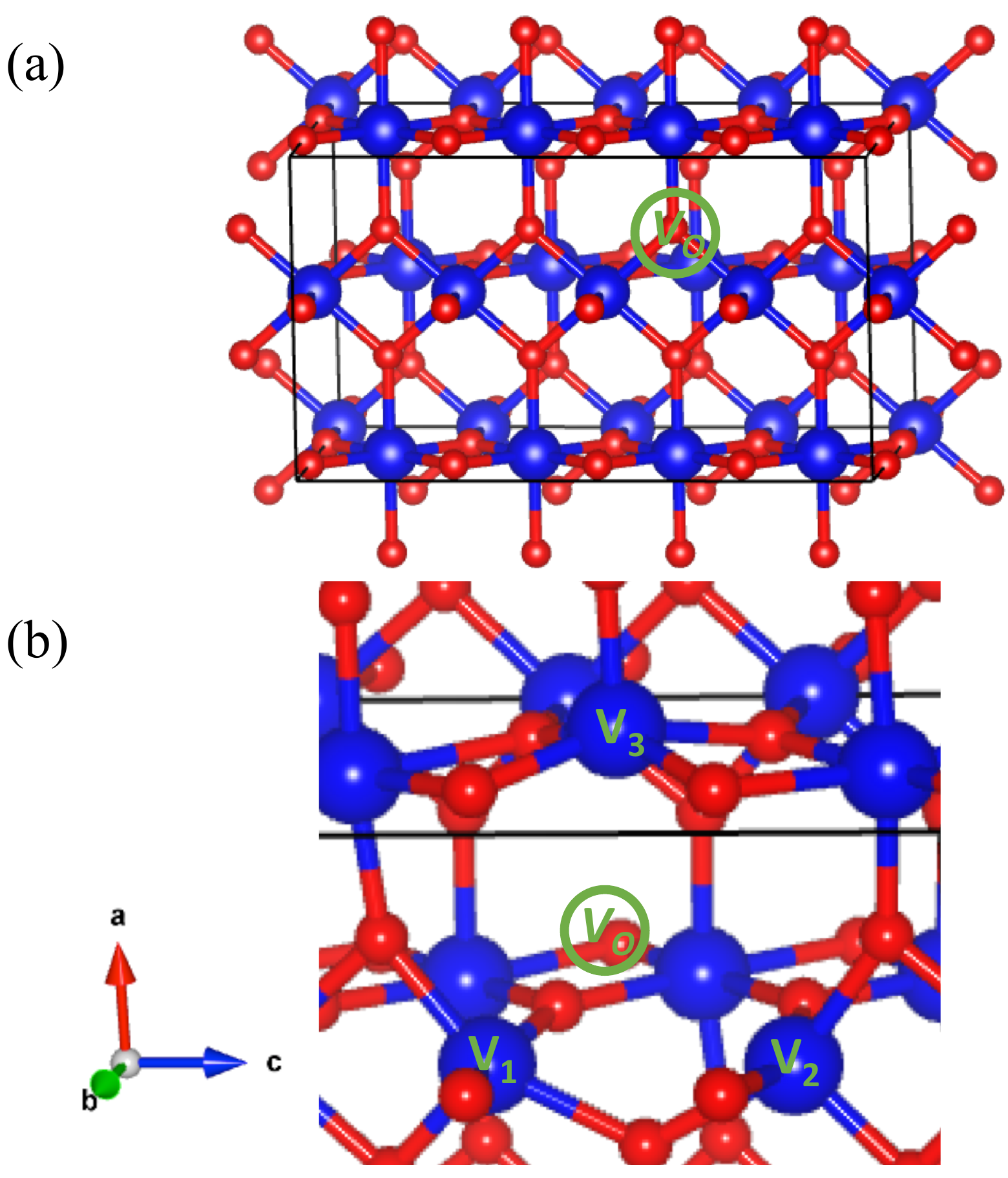}
 \caption{(a) shows the DFT-SCAN relaxed structure of our 48-atom \van~supercell. The \vac~site is labeled. (b) shows the strong local distortions seen around the \vac~site in relaxed VO$_{2-\delta}$.  The nearest neighbor V-atoms to the \vac~site are also labeled. (V(blue), O(red))}
 \label{fig:struct}
\end{figure}

$\sl Introduction$: Defects determine and control properties of solids and to a large degree impart specific functionalities to them.~\cite{vanDeWalle_RPM} Harvesting this will play a key role in advanced electronic materials for future information technologies, such as neuromorphic~\cite{StanWilliams_Neuromorphic} and quantum computing.~\cite{vanDeWalle_QuantComp} There is a significant gap in our understanding of how defects influence technologically relevant phase-transitions, such as the metal-insulator transition (MIT), in strongly correlated materials~\cite{Mottronics_review,Mottronics_VO2}. Closing this gap is particularly challenging when strong non-local electron-electron correlation effects are coupled with the strong local interactions of defects with the lattice~\cite{Ganesh_VO2,Polaron,Zylberstejn}; such a fundamental understanding is necessary for paving the way for robust future technologies based on correlated materials.

\van~is formally a 3$d^1$ system which is expected to have a metallic ground-state owing to its half-filled $d$-band. While it is a metal at high temperatures, it is regarded as a ``bad metal" because its resistivity is above the Mott-Ioffe-Regel bound~\cite{BadMetals_Emery,Science_VO2_BadMetal}, and as such it is a strongly-correlated material. Below the transition temperature --  T$_{MIT} \sim$ 340~K, \van~becomes insulating.~\cite{Goodenough,Sawatzky_PES} This electronic transition is accompanied by a structural phase transition, in which the high-temperature and high-symmetry rutile ($ \sl R$) phase [Fig.~\ref{fig:struct}(a)] transforms to a low-temperature low-symmetry monoclinic ($\sl M$) phase by the formation of V-V dimers along the rutile $\bf{c}$ axis. As such this MIT is considered a Peierls-Mott type transition.~\cite{DMFT_Biermann_vo2,DMFT_Brito_vo2}. It has been a long-standing problem to understand if the MIT is driven primarily by electron correlations~\cite{Parkin_vo2_prl2016} or by intrinsic structural instabilities~\cite{anharmonic}, and understanding this is crucial to control MIT in correlated solids. 

The nominally V$^{4+}$ vanadium atoms in the rutile phase sit at the center of distorted oxygen octahedra. This splits the otherwise low-energy triply degenerate $t_{2g}$ manifold of V-3$d$ into a single $a_{1g}$ orbital and a doubly degenerate $e_{g}^{\pi}$ combination of orbitals. It is the $a_{1g}$ orbital in the rutile phase that takes part in forming the dimerized singlet state across the MIT, thereby opening up an electronic gap. Indeed, detailed X-ray absorption, X-ray diffraction and transport studies on thin films across the MIT reveal tunability of $T_{MIT}$ with strain, which appears to correlate with changes in $a_{1g}$ orbital occupancy.~\cite{Parkin_vo2_NPhys2013,Parkin_vo2_prl2016}

Doping \van~with external substitutional dopants such as W in the V site has been shown to tune $T_{MIT}$, and partially reverts it back to normal behavior.~\cite{Science_VO2_BadMetal} Doping thin films of \van~with intrinsic oxygen vacancies was recently shown to fully suppress the metal to insulator transition with no observed structural transitions down to 50~K.~\cite{Nonstoich_VO2} It is not clear what is the underlying mechanism behind the tunability or suppression of the MIT with doping, and how this tunability/suppression depends on the doping concentration or the type of dopant. Understanding this can perhaps also reveal what factors are crucial in driving the MIT in the first place in stoichiometric \van.

In this Letter, we investigate the influence of doping \van~with \vac~and both capture and explain the experimentally observed suppression of the Peierls-Mott type MIT. We accomplish this by using density functional theory based methods (DFT and DFT+U), together with correlated-electronic structure methods, such as dynamical mean-field theory (DMFT)~\cite{DMFT_Frank_rutile} and quantum Monte Carlo (QMC)~\cite{Wagner}. QMC is a very accurate ground state correlated-method which gives accurate vacancy formation energies and electron densities. The use of DMFT allows us to expand the investigation into excited state spectra. We obtain critical insights on what factors are crucial in driving and controlling the MIT in \van, which should be relevant for similar bad-metal systems.~\cite{LMO_MIT,Parkin-SrCoO3-x,Bad-Metal-Quant-Critical} We also assess the importance of including electronic correlations in describing the physics of doping with defects and elucidate the link between correlations and orbital filling. We conclude that electronic interactions are the primary driver of the MIT, and the structural distortion is a secondary consequence.

$\sl Methods$: Atomic structures were relaxed using atomic forces obtained from Density Functional Theory (DFT) based calculations with the SCAN functional~\cite{SCAN_PRL,SCAN_NatChem} [comparable results were obtained with the DFT+U approach as detailed in the Supplementary Information (SI)]. All calculations were performed in a 48-atom supercell as shown in Fig.~\ref{fig:struct}(a). Oxygen vacancy structures were obtained by removing one oxygen atom from the 48-atom supercell (corresponding to VO$_{2-\delta}$ with $\delta$=0.0625). The relaxed structure (shown in Fig.~\ref{fig:struct}(b)) indicates a significant degree of local distortion around the \vac~site. It is not immediately clear how important these distortions are in describing the observed suppression of MIT. To assess this, we also performed calculations using a virtual crystal approximation (VCA) method whereby electrons are added to \van~to mimic electronic-doping, without any atomic relaxation/distortion. Calculations beyond DFT, such as a charge self-consistent DFT+DMFT method~\cite{gri12} using the TRIQS package~\cite{par15, set16}
as well as the Diffusion Monte Carlo (DMC) flavor of continuum QMC method using QMCPACK\cite{kim_qmcpack_2018, kylanpaa2017, Ganesh_VO2} were employed to account for strong electronic correlations on the V sites. Further details about these methods are provided in the SI.

\begin{figure}[b]
 \includegraphics[width=3.3in]{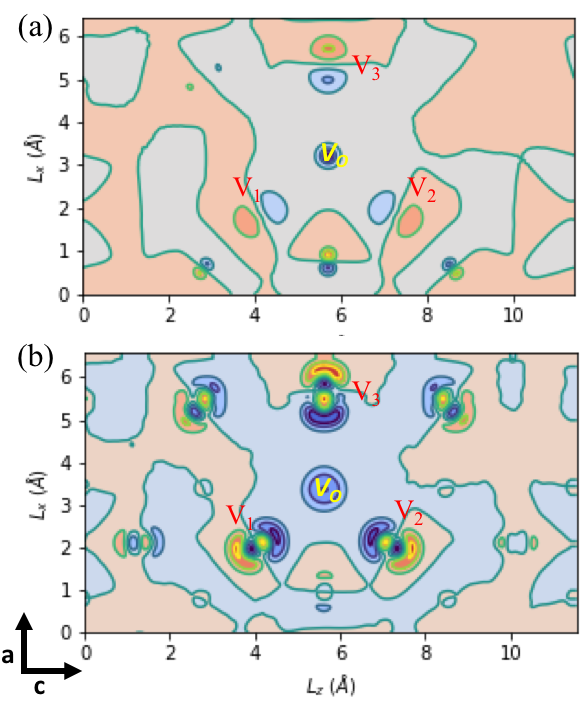}
 \caption{Contour plots of charge-density difference (units of $\sl{e}/(a.u.)^3$) between non-stoichiometric and stoichiometric \van~in the central plane containing the \vac~from (a) DMFT and (b) QMC calculations. Both levels of theory suggest that electronic reorganization propagates from local changes around the defect site. The three V-atoms closest to the \vac~site [labeled in Fig.~\ref{fig:struct}(b)], show the largest density changes. (blue/red show density loss/gain compared to pristine \van. The color-bar ranges from $\{-4,3\}$ and $\{-1.2,1.2\}$ for DMFT and QMC plots, respectively.)}
 \label{fig:chgdens}
\end{figure}

\begin{figure*}[htp]
 \includegraphics[width=5.0in]{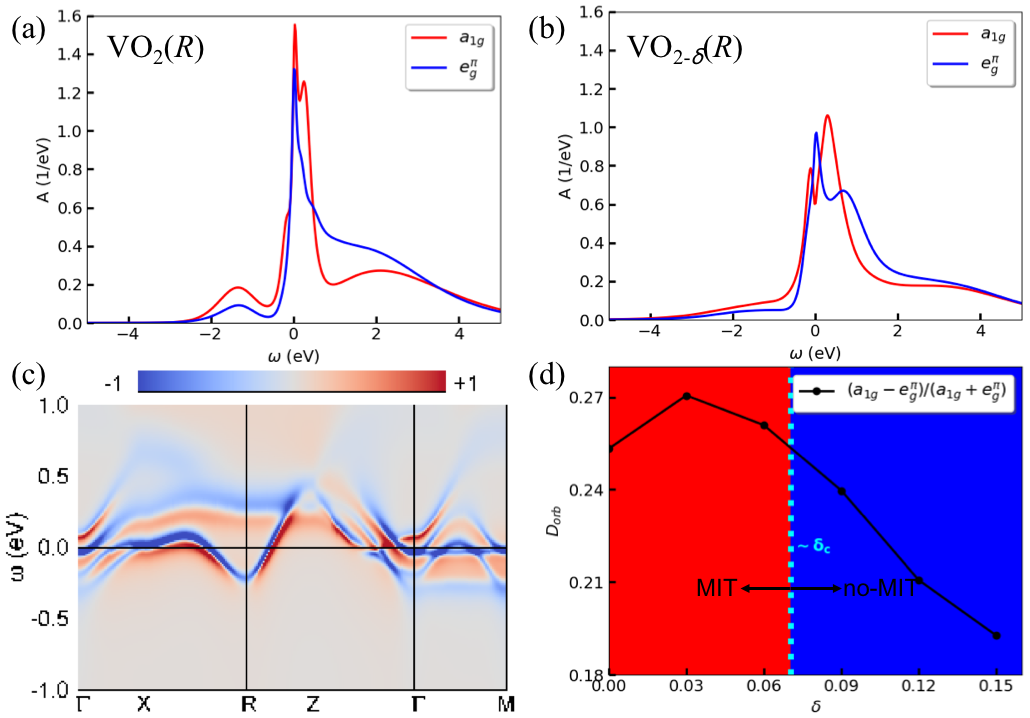}
 \caption{(a) and (b) show the orbitally-resolved local V-$t_{2g}$ spectral functions $A(\omega)$ obtained from DMFT calculations for pristine (\van) and non-stoichiometric (VO$_{2-\delta}$) $\sl R$-phase, respectively. In the pristine case, a lower Hubbard peak at $\sim 1.35$ eV is seen that is predominantly of $a_{1g}$ type, indicative of strong electronic correlations. Introducing \vac's~suppresses both the quasiparticle peak at low energy as well as the Hubbard peak. (c) shows the difference of the $\bf k$-resolved  spectral function between defective and pristine \van~within VCA, for a charge-doping of 0.06 $e^-$. Significant reorganization of spectral weights is observed, with a stronger reduction around $\omega=0$. (d) Orbital-dichroism ($D_{orb}$) from DMFT-VCA calculations shows that with electron-doping the dichroism inititially increases, but subsequently becomes less positive with a diminished magnitude. At $\delta_c\sim 0.07$ its value dips below that of pristine \van, and is indicative of a doping-level beyond which MIT would be suppressed. This value is close to the experimental estimation of $\delta_{c}^{exp}$=0.098 in \van~thin-films.}
 \label{fig:DMFTSpecF}
\end{figure*}

\begin{em}
Results and Discussion:
\end{em}The \vac~formation energies (defined in the SI) obtained from DMFT and QMC-calculations are 5.09 and 5.10(1)~eV, respectively, in excellent agreement with each other. As such both of these strongly-correlated techniques must capture the changes in the underlying electronic structure due to \vac's appreciably well, in turn implying a coupling of \vac's to the strength of electron correlations. The charge density difference compared to stoichiometric \van~from DMFT and DMC results are shown in Fig.~\ref{fig:chgdens} (a) and (b), respectively. While absolute values are not identical, possibly due to different types of pseudopotentials (see SI),  at both levels of theory, the largest change in charge density appears to be around the three vanadium atoms closest to the \vac~as shown in Fig.~\ref{fig:struct}(b). Specifically, QMC charge densities show that the three underbonded V$_{1,2,3}$ atoms have an excess of electrons compared to pristine \van~(a cutoff distance of 1.26~\AA~was used for charge integration). A local perturbation of the charge density in the presence of \vac, as opposed to complete delocalization of the excess electrons from \vac~as one might expect in a regular metal, could be reminiscent of its bad-metal behavior. Also, the agreement between DMFT and DMC on ground-state properties such as vacancy formation energy and charge-density gives greater confidence in the DMFT excited-state properties, such as the spectral function. 

Fig.~\ref{fig:DMFTSpecF}(a) shows the orbitally-resolved local spectral function $A(\omega$) of V-3$d^1$ for stoichiometric $R$-phase \van~obtained at a temperature of $T$ = 370~K, well above T$_{MIT}$ = 340~K. In addition to a narrow quasiparticle peak close to $\omega$ = 0, a lower Hubbard peak is seen at $\sim$1.35~eV below the Fermi-level. This is in contrast to the density-of-states obtained from DFT-based methods (see SI), but in good agreement with the photoelectron spectroscopy (PES) results~\cite{Sawatzky_PES} that show a small low-energy bump in the valence spectrum of \van~around $\sim 1$~eV. We find these states to be dominated by $a_{1g}$ orbital contributions compared to the two degenerate $e_{g}^{\pi}$ orbitals (only the average $e_{g}^{\pi}$ contribution is shown). The same PES experiment~\cite{Sawatzky_PES} shows these localized V-3$d^1$ states to be occupied even in the insulating $M$-phase, but they are now part of the doubly-occupied V-V dimerized singlet state, which is also composed of the $a_{1g}$ orbitals. Therefore one might expect that disrupting this near orbital polarization of $a_{1g}$ orbitals should have an influence on the MIT in \van. Indeed, X-ray dichroism experiments~\cite{Parkin_vo2_NPhys2013} showed that lowering orbital polarization of $a_{1g}$ via strain led to a reduced $T_{MIT}$. We now explore the influence of \vac's on orbital occupancies and the resultant orbital dichroism, defined as $D_{orb}=(a_{1g}-e_{g}^{\pi})/(a_{1g}+e_{g}^{\pi})$, which loosely corresponds to the measured quantity in  X-ray dichroism experiments~\cite{Parkin_vo2_NPhys2013,Parkin_vo2_prl2016}.

The orbitally-resolved spectral function from DMFT for VO$_{2-\delta}$ [Fig.~\ref{fig:DMFTSpecF}(b)] is very different from that of \van. Clearly, the low-energy bump at $\sim$ 1.35~eV has a much-reduced intensity, and the width of the quasiparticle peak has significantly widened, leading to its reduced peak height -- suggesting weakened electron-electron correlations. In particular, the $e_{g}^{\pi}$ orbital occupancy appears to be competing with the $a_{1g}$ occupancy, indicating that the near orbital $a_{1g}$ polarization seen in \van~is now absent. The integrated V-$d$ charge is now 1.12$e^-$, much different from the nominal 1$e^-$ in stoichiometric \van, and consistent with the QMC charge density analysis discussed above. This strongly indicates that in the presence of \vac, reduction of the near orbital polarization of the $a_{1g}$ orbital and concomitant weakening of electronic correlations lead to suppression of the MIT. Note furthermore that in contrast to \vac's in, e.g., band-insulating SrTiO$_3$, no straightforward $e_g$-like (localized) in-gap states are detected.

In order to investigate the importance of capturing the large local distortion seen in our DFT-relaxed structure [Fig.~\ref{fig:struct}(b)] we performed DMFT calculations in the \van~unit-cell using a virtual-crystal approximation (VCA). The VCA calculations were performed with 0.06 more pseudo charge on oxygen, resulting in a 1.12 electron filling of the V site as seen in the explicit \vac~DMFT-calculation. The difference plot of the quasiparticle dispersion as shown in Fig.~\ref{fig:DMFTSpecF}(d) is consistent with the changes observed in the explicit spectral function. Significant reorganization of spectral weights is observed with doping, with a stronger reduction around $\omega=0$. This suggests that while details of the distortion are necessarily important in getting the correct vacancy-formation energetics as well as correlating local bond-disproportionation around the defect to charge-reorganization to obtain the correct orbital weights, qualitative changes in the relative orbital occupancies with doping should be captured even in the absence of such localized distortions.

Suppression of the MIT in non-stoichiometric \van~is consistent with recent experiments where the metallic $R$-phase was seen to be stable even down to 50~K~\cite{Nonstoich_VO2}. Since this behavior is experimentally similar to $W$-doped \van, it appears that introducing \vac~in \van~is similar to electronically doping it, even in the metallic phase, as confirmed by DMFT.  Given that the changes are local, as inferred from our charge-density plots (Fig.~\ref{fig:chgdens}), our calculations suggest that there is some kind of a percolation threshold that needs to be exceeded in order to suppress the MIT. This hypothesis is in agreement with experimental observations which estimate a critical doping concentration of $\delta_{c}^{exp} = 0.098$ to be necessary to fully suppress the metal-insulator transition. 

\begin{figure}[b]
 \includegraphics[width=3.3in]{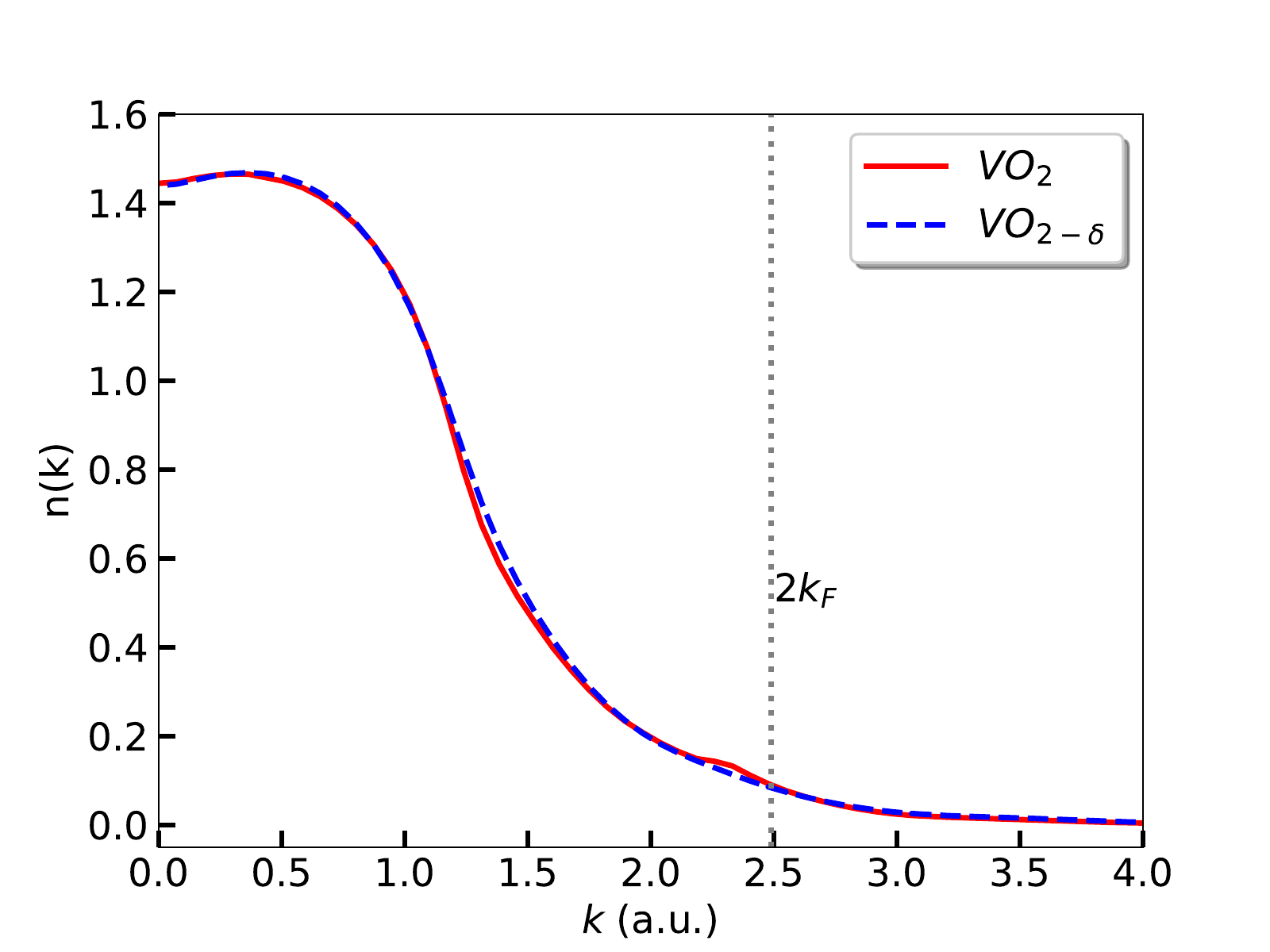}
 \caption{Momentum density along $k$ direction conjugate to the V-V dimerization direction for pristine and non-stoichiometric \van~from LSDA+U calculations. Vanishing of the $\sim 2k_F$ peak in VO$_{2-\delta}$ suggests a lack of a charge-density wave (CDW) when \vac's are introduced, thereby altering its bad-metal behavior.}
 \label{fig:momdens}
\end{figure}

To see the presence of such a critical concentration, and its relationship as well as manifestation in terms of the suppression of the near orbital-polarization, we estimate $D_{orb}$ as defined above from VCA calculations performed at a series of doping concentrations ($\delta$) as shown in Fig.~\ref{fig:DMFTSpecF}(d). $D_{orb}$ becomes initially more positive with $\delta$, but quickly reduces in magnitude, consistent with the expected relative loss of the $a_{1g}$ orbital-weights. Above $\delta = \delta_{c}\sim0.07$, $D_{orb}$ falls below its pristine \van~value, indicative of a critical concentration above which no MIT can be realized. This observation proves the direct link between suppression of near-orbital polarization of $a_{1g}$ and the suppression of the MIT. Given that $D_{orb}$ can be loosely connected to the experimentally measured X-ray dichroism~\cite{Parkin_vo2_NPhys2013}, our results suggest that the physics underlying MIT suppression and the tuning of T$_{MIT}$ in \van~by means of \vac's, extrinsic-dopants, strain, or electric fields is very similar, and all of these approaches are different controls to modify the $a_{1g}$'s relative occupancy and reduce electron-electron correlations to influence the MIT. X-ray dichroism studies in well-controlled non-stoichiometric \van~thin-films should experimentally confirm this claim. 

Electron-electron correlations are thought to be important for describing the bad-metal behavior of \van~ 
as well as causing V-V dimerization in the $M$ phase.~\cite{DMFT_Brito_vo2, Parkin_vo2_prl2016} 
A 1-D Peierls instability could be driven by a 1-D charge-density wave (CDW) instability, which would correspond to a nesting vector at $k \sim 2k_F$, where $k_F$ is the Fermi wave-vector. If this is indeed the case, it will show up as a `bump' in the momentum distribution $n(k)$. When using a value of `U' optimized by DMC, the $n(k)$ from LSDA+U and DMC have negligible differences for $k > k_F$~\cite{Illkka_momdens}, as such we study $n(k)$ using LSDA+U (details in SI). Indeed, the $n(k)$ obtained from our LSDA+U calculation for the $R$-phase \van~shows such a bump at $ \sim 2k_F = 2.49$~a.u., where the $k$ vector is plotted conjugate to the ${\bf c}$ direction as shown in Fig.~\ref{fig:momdens} and recently published by some of us in Ref.~\cite{Illkka_momdens} No such oscillatory features are observed along the other $k$-directions. This shows the presence of a 1-D CDW instability with a nesting vector of $2k_F$, which would be required to drive the V-V dimerization, resulting in the insulating $M$-phase. The bump is also expected to cause divergences in the response-functions, such as the Lindhard-susceptibility, and thereby lead to its bad-metal behavior -- i.e. the violation of the Wiedemann-Franz law. 

In the presence of \vac, this bump at $ \sim 2k_F$ vanishes (Fig.~\ref{fig:momdens}), altering the bad-metal characteristics of \van, and the density monotonically decreases with increasing momentum. This would result in divergence-less response functions, and thereby a lack of an electronic instability to drive a phase-transition. This is a consequence of the weakening of electron-electron correlation, as also suggested from our DMFT results, whereby the filling is not 1$e^-$ on each V site and
the near orbital polarization of $a_{1g}$ is lacking. Suppressing this electronic driving force for a Peierls distortion should also result in suppression of the structural transition to the $M$-phase.

$\sl Conclusion$:We conclude that oxygen-vacancies donate electrons, and the resulting charge-doping reduces the near orbital polarization of the $a_{1g}$ orbitals seen in R-phase \van. This reduction is associated with a weakening of the electronic correlations and leads to a reduced drive for V-V dimerization. This not only  suppresses the MIT, but also substantially alters the ``bad metal" characteristics of the metallic phase in non-stoichiometric \van. Our study re-affirms the intricate connection between structural dimerization and electron-correlation in \van, and suggests that the MIT is predominantly driven by a correlation-induced electronic instability and not a structural instability, with the $a_{1g}$ occupancy being the primary knob to control the MIT via different external perturbations such as \vac's, extrinsic dopants, strain, or electric-fields. This understanding of the presence of a single fundamental knob that allows control over complex coupled phase-transitions in correlated solids can be relevant to engineering functional interfaces of correlated oxides and also aid in understanding technologically relevant MIT systems that show concomitant magnetic- and/or structural-transitions, such as manganites~\cite{LMO_MIT} and the family of non-stoichiometric ABO$_{3-x}$ compounds.~\cite{Parkin-SrCoO3-x}.

$\sl Acknowledgement$: This work was supported by the U.S. Department of Energy, Office of Science, Basic Energy Sciences, Materials Sciences
and Engineering Division, as part of the Computational Materials
Sciences Program and Center for Predictive Simulation
of Functional Materials. This research used resources of the
Oak Ridge Leadership Computing Facility at the Oak Ridge
National Laboratory, which is supported by the Office of
Science of the U.S. Department of Energy under Contract
No. DE-AC05-00OR22725. This research used resources of
the National Energy Research Scientific Computing Center,
a DOE Office of Science User Facility supported by the
Office of Science of the U.S. Department of Energy under
Contract No. DE-AC02-05CH11231. This manuscript has
been authored by UT-Battelle, LLC under Contract No.
DE-AC05-00OR22725 with the U.S. Department of Energy.
The United States Government retains and the publisher, by
accepting the article for publication, acknowledges that the
United States Government retains a nonexclusive, paid-up,
irrevocable, worldwide license to publish or reproduce the
published form of this manuscript, or allow others to do so,
for United States Government purposes. The Department of
Energy will provide public access to these results of federally
sponsored research in accordance with the DOE Public Access
Plan (http://energy.gov/downloads/doe-public-access-plan)

FL acknowledges financial support from the DFG project LE 2446/4-1. 
DFT+DMFT computations were performed at the JURECA Cluster of the J\"{u}lich Supercomputing Centre (JSC) under project
number hhh08.

\newpage

{\bf Supplementary Information:} 
\maketitle

\begin{figure*}[htp]
 \includegraphics[width=5.5in]{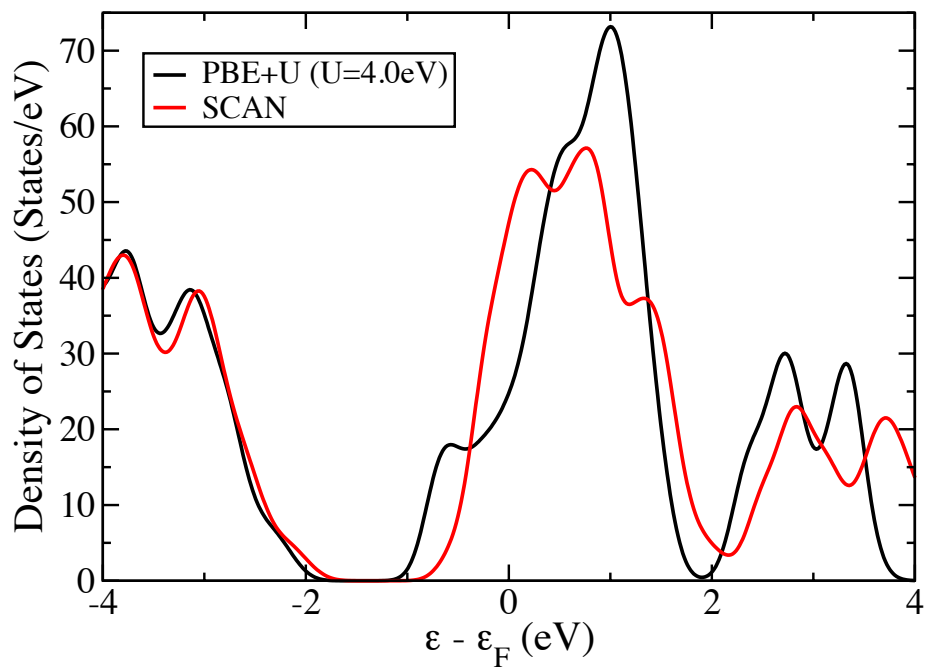}
 \caption{Total density-of-states for \van~from SCAN and PBE+U calculations.
 }
 \label{fig:DFT-DOS-Stoich}
\end{figure*}

\begin{figure*}[htp]
 \includegraphics[width=5.5in]{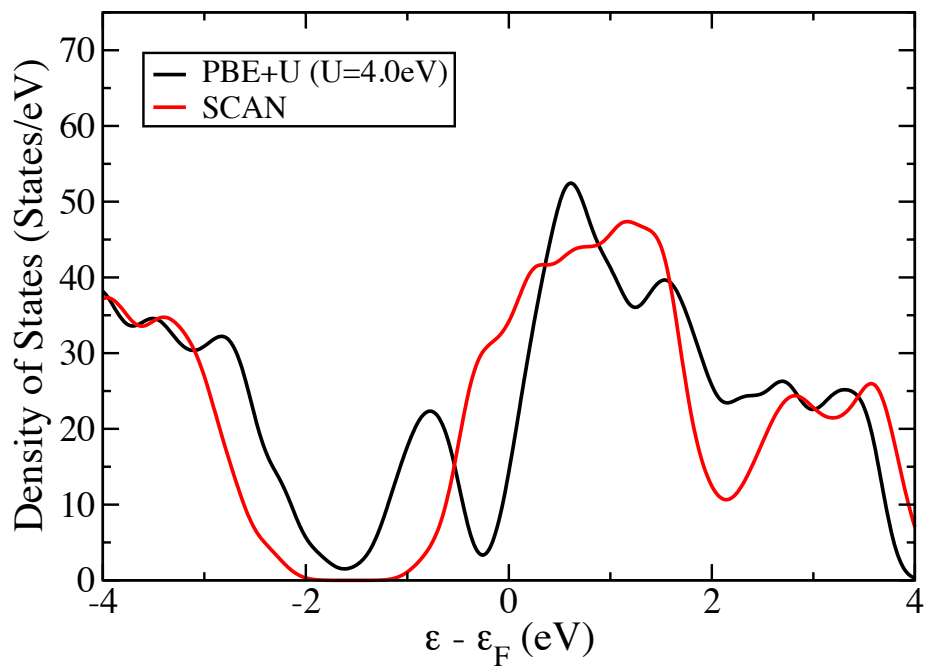}
 \caption{Total density-of-states for non-stoichiometric \van~from SCAN and PBE+U calculations.
 }
 \label{fig:DFT-DOS-NonStoich}
\end{figure*}

Atomic structures were relaxed using atomic forces obtained from non-spin polarized Density Functional Theory (DFT) based calculations with the SCAN meta-generalized-gradient (meta-GGA) functional~\cite{SCAN_PRL,SCAN_NatChem} as well as the Perdew, Burke, and Ernzerhof (PBE) functional with a Hubbard `U' i.e. PBE+U, with U=4eV in the rotationally invariant Dudarev~\cite{Dudarev} approach, as implemented in the VASP package~\cite{Kresse1} using PAW pseudopotentials~\cite{Kresse2}. All calculations were performed in a 48-atom supercell with dimensions: $L_x$ = $L_y$ = 6.4412 \AA~and $L_z$ = 11.4112 \AA, along the orthogonal $\bf{a,b,c}$ directions as shown in Fig.1(a) in the main-text. Atomic relaxations were performed using a 4x4x4 {\bf k}-mesh to perform the Brillouin-zone integration with a gaussian-smearing of 0.2 and a plane-wave kinetic-energy cutoff of 400 eV, using the  'accurate' setting for the precision tag in VASP. The 'V' and 'O' PAW potentials had 13 and 6 valence electrons, respectively. Forces were converged down to 0.01 eV/\AA. The supercell  Oxygen vacancy structures were obtained by removing one oxygen-atom from the 48-atom supercell (corresponding to VO$_{2-\delta}$ with $\delta$=0.0625). Relaxed structure (shown in Fig.1(b) in the main-text) suggests a significant degree of local distortion around the \vac~site. It is not immediately clear how important are these distortions in describing the observed suppression of MIT. To assess this, we also perform calculations using a virtual crystal approximation (VCA) method whereby electrons are added to \van~to mimic electronic-doping, without any atomic relaxation/distortion. In more detail, an oxygen-like pseudo-atom of charge $Z=8+\delta$, i.e. a nominal mixture of an O and an F atom, is utilized within VCA to mimic the appearance of oxygen vacancies. The SCAN and PBE+U total density-of-states (DOS) for pure and non-stoichiometric \van~are shown in Fig.~\ref{fig:DFT-DOS-Stoich} and~\ref{fig:DFT-DOS-NonStoich}, respectively. 

Calculations beyond DFT are put into practice to account for strong electron correlations on the V sites. We use a charge self-consistent DFT+dynamical mean-field theory (DMFT) framework~\cite{gri12}, building up on a mixed-basis pseudopotential approach for the DFT part and the continuous-time quantum-Monte-Carlo method, as implemented in the TRIQS 
package~\cite{par15, set16}, for the DMFT impurity problem. The GGA in the PBE-functional form is employed within the Kohn-Sham cycle. Vanadium 3d4s4p and Oxygen 2s2p were treated as valence electrons in the pseudopotential generation scheme.  
Locally, threefold effective V(3$d$) Wannier-like functions define the correlated subspace, which as a whole consists of the corresponding sum over the various V sites in the defect problem. Projected-local orbitals~\cite{ama08} of 3$d$ character provide the effective functions from acting on Kohn-Sham conduction states above the O(2$p$)-dominated band manifold. The selected threefold functions are given by the local three-orbital sector lowest in energy, respectively. Each V site marks an impurity problem, and the number of symmetry-inequivalent vanadium sites provides the number of single-site DMFT problems to solve. A three-orbital Hubbard Hamiltonian of Slater-Kanamori form, parametrized by the Hubbard $U$ = 4 eV and the Hund’s exchange $J_H$ = 0.7 eV, acts on each V site. These values for the local Coulomb interactions are close to an effective $U_{eff} = U - J$=3.5 eV, which was shown to be optimal in previous studies on \van.~\cite{kylanpaa2017,Ganesh_VO2} A double-counting correction of the fully-localized form~\cite{ani93} is utilized. The analytical continuation of the finite-temperature Green’s functions on the Matsubara axis $i\omega$ to real frequencies is performed via the maximum-entropy method
as well as with the Pad{\'e} scheme. The $\bf{k}$-resolved DMFT spectra for pristine and defective \van~is shown below in Fig.\ref{fig:DMFT_spectra}.

The Diffusion Monte Carlo (DMC) flavor of continuum quantum Monte Carlo methods was performed on the same structure using QMCPACK\cite{kim_qmcpack_2018}. Diffusion Monte Carlo is a highly accurate wavefunction based projector method that improves variationally as the starting, or trial, wavefunction is improved.  An accurate trial wavefunction is essential to minimize residual fixed node/phase error in the method.  The basic form of the trial wavefunction used here is a product of an up/down spin factorized Slater determinant and a Jastrow correlation factor, as follows:
\begin{align}
	\Psi_T(R) = e^{J(R)}D^\uparrow(R^\uparrow)D^\downarrow(R^\downarrow).
\end{align}
The nodal/phase structure of the trial wavefunction is determined by the single particle orbitals populating the determinants.  Trial orbitals were obtained within LSDA+U via the Quantum Espresso code for all atomic structures.  The U value was selected to be 3.5 eV since this value minimizes the variational DMC total energy for \van~as demonstrated by prior studies~\cite{kylanpaa2017,Ganesh_VO2}. A ferromagnetic configuration was chosen for the V-sites since this arrangement of magnetic moments is more robust to changes in the lattice induced by defects.  Since the spin gap in the materials is small, this choice has a negligible impact on the resulting defect formation energies.  The trial Jastrow factor was represented as a sum of one- and two-body correlation factors represented in a B-spline basis along electron-electron or electron-ion pair distances.  The Jastrow factor was variationally optimized with respect to the total energy using the linearized optimization method.  The bulk optimized Jastrow factor was used in both bulk and defective phases to minimize the potential impact of pseudopotential locality errors in the subsequent DMC calculations, similar to what is commonly done for van der Waals systems.  In both the bulk and defective cases, the absolute variance to energy ratio resulting from the Jastrow was near $0.026$ Ha, indicating uniform quality across the structures.  Diffusion Monte Carlo total energies and spin densities were obtained by averaging over a 2x2x2 supercell twist grid.  DMC runs at each twist were performed with a large population of random walkers ($\approx 14,000$ walkers per twist) and a small timestep of $0.005$ Ha$^{-1}$ resulting in an acceptance ratio of 99.6\%.  Validated\cite{krogel2016_2,dzubak2017,kylanpaa2017} norm-conserving RRKJ pseudopotentials were used for vanadium (Ne-core) and oxygen (He-core).  The T-move scheme was used to maintain the variational principle in DMC calculations involving these non-local pseudopotentials. All QMC related simulation workflows were driven with the Nexus\cite{krogel2016} workflow automation system. 

Since the oxygen dimer formation energy has different amounts of error in the different methods used, we calculate the oxygen vacancy formation energy in \van using atomic oxygen and bulk \van ~as reference using this formula: 
\[
E_f[V_O] = E_{\rm total}[n({\rm VO}_{2-\delta})] - E_{\rm total}[n({\rm VO}_2)] - E[{\rm O}],
\]
where the total energies are calculated at 0K and $n=16$ is the number of formula units in our 48-atom supercell, and $\delta = 0.0625$. $E[{\rm O}]$ is taken to be one-half the total energy of an oxygen dimer.

\begin{figure*}[htp]
 \includegraphics[width=5.5in]{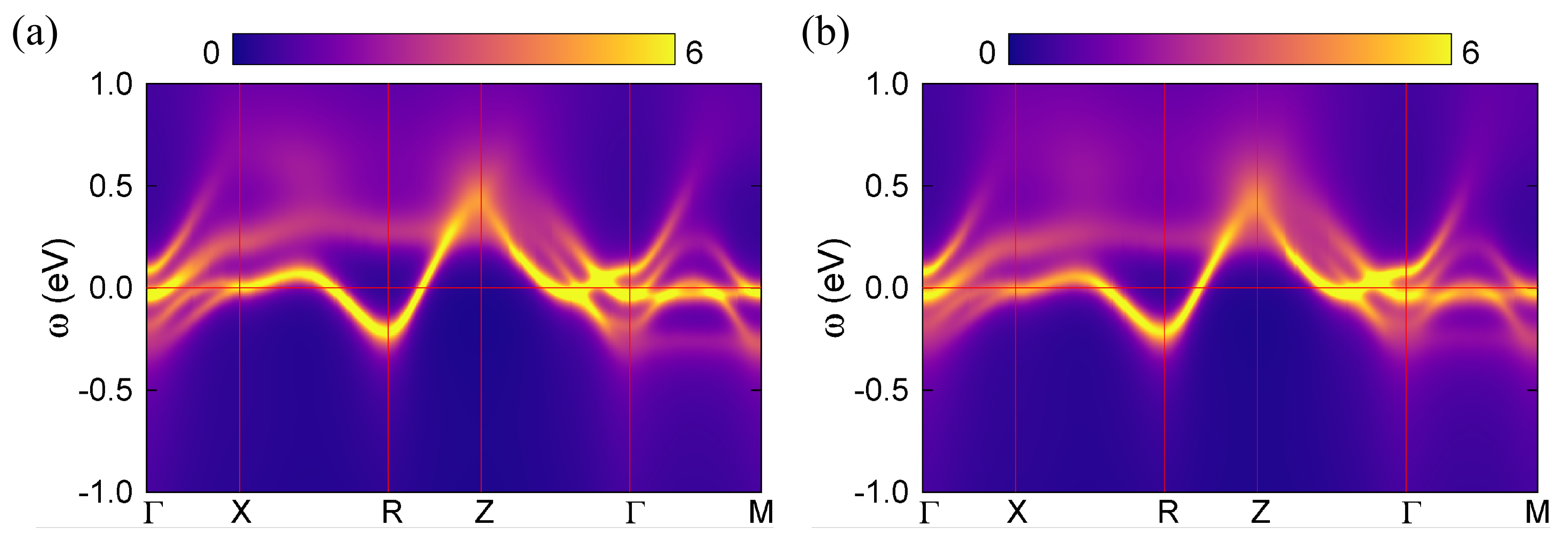}
 \caption{(a) and (b) $k$-resolved  spectral function of pristine \van~and VO$_{2-\delta}$ (within VCA) for $\delta \sim$ 0.06 $e^-$, respectively from our DMFT calculations at $T \sim$ 370K.
 }
 \label{fig:DMFT_spectra}
\end{figure*}

\bibliography{vo2}
\end{document}